\documentstyle[twoside, epsfig]{article}
\input ibvs2.sty

\begin{document}

\IBVShead{0000}{16 September 2011}

\IBVStitle{V1162 Orion: Updated Amplitude and Period Variation}

\IBVSauth{Khokhuntod, Pongsak$^{1,2}$; Zhou, Ai-Ying$^{3*}$;
Boonyarak, Chayan$^{1}$; Jiang, Shi-Yang$^{1,3}$}

\IBVSinst{Department of Physics, Faculty of Science,
Naresuan University, Phitsanulok 65000, Thailand}

\IBVSinst{Department of Astronomy, Beijing Normal University, China}

\IBVSinst{National Astronomical Observatory of China,
Chinese Academy of Sciences, Datun Road 20A, Beijing, 100086,
China; *Correspondent author, E-mail: {\tt aiying@nao.cas.cn}}

\IBVStyp{DSCT}

\IBVSkey{CCD photometry}

\IBVSabs{In 2010/11/23-25, we find the amplitude of V1162 Orion, }
\IBVSabs{ a $\delta$ Scuti Stars, in V recovered to about 0.18 mag.}

\begintext


V1162 Orion is an intermediate-amplitude $\delta$ Scuti type variable.
Its variability was first discovered by Lampens (1985),
who derived a period of 0.078686(2) days with a full amplitude of
about 0.18 mag in $V$ band.
Hintz et al. (1998) observed this star and found its amplitude in $V$
had dropped to about 0.10 mag.
Arentoft \& Sterken (2000) and Arentoft et al. (2001a, 2001b)
got the same amplitude of $\sim$0.10 mag in $V$.

We had observed this star from 2007 January  to 2010 February
using several telescopes at Naresuan University, Thailand and at Yunnan Observatory and
the Xinglong Station of National Astronomical Observatories, China (NAOC).
We determined a sum of 39 new times of maximum light
(refers to Table\,\ref{maximum}), following the method described by Zhou \& Liu (2003).
Errors involved in the maximum determination are around 0.00045\,d or less.
In this note, we present a two-night sample of the newly observed light curves
along with an updated $O-C$ diagram.
Figure~1 shows the light curve of V1162 Ori on 2010 November 24 and 25
from the 60-cm telescope of NAOC.
We can see clearly that the peak-to-peak full amplitude in $V$ recovered to about 0.18\,mag.
This conforms with the amplitude variability declared in the literature.
The differential photometric light curve data
are available upon request from the authors.


\begin{figure*}[tbh]
\centering
\includegraphics[angle=-90,width=0.94\columnwidth]{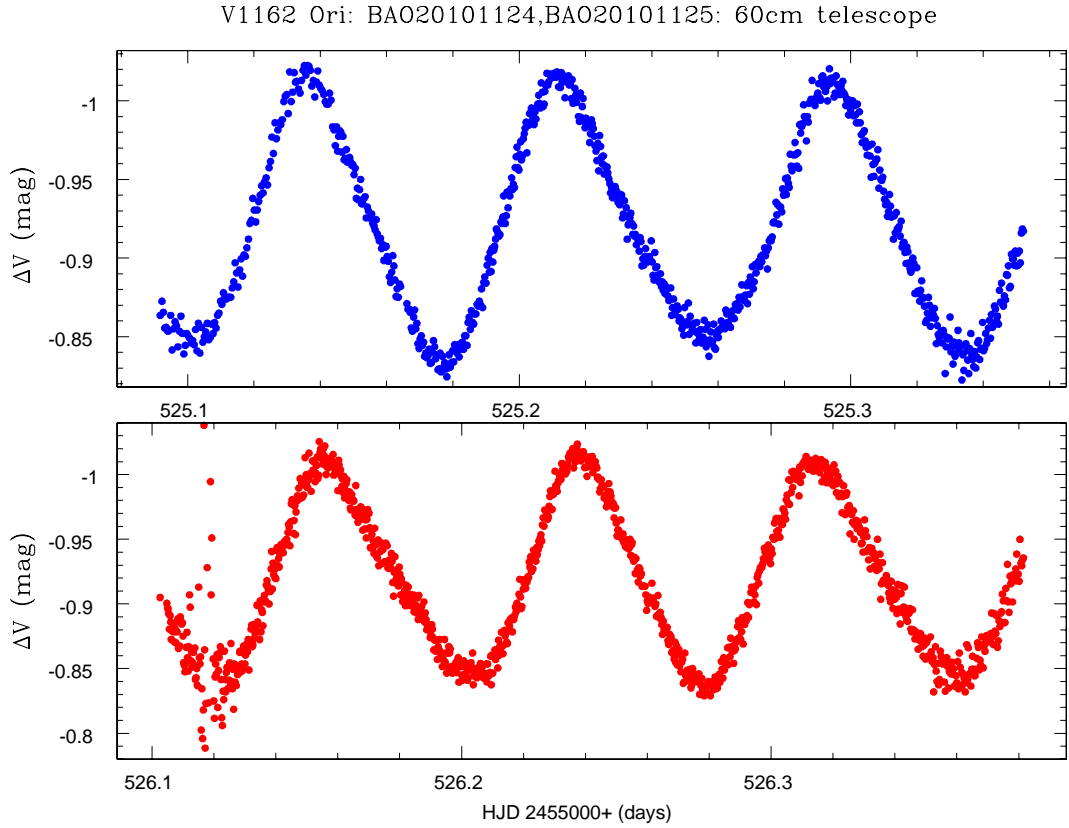}
\caption{The light curves of V1162 Ori obtained
on 2010 Nov. 24 and 25.}
\label{fig:LC}
\end{figure*}

We have noted that Arentoft et al.\,(2001a) presented the results of multisite observations
contributed by 18 telescopes from 15 observatories around the global.
They measured 253 light extrema (145 maxima and 108 minima) during the observing season from
1999 October to 2000 May.
They got an overall tendency of period decreasing
with a very high rate of $\frac{{\rm d}P}{{\rm d}t}$=$-5\times 10^{-9}$ s s$^{-1}$
(equivalent to $\frac{1}{P}\frac{{\rm d}P}{{\rm d}t}=-2.4\times10^{-5}$ yr$^{-1}$).
Moreover, the $(O-C)$ residuals seemed to follow a cyclic variation with
a period of 285$\pm$3\,d,
which is on a time-scale similar to that of the amplitude variations (282$\pm$6\,d).
In addition, we found 40 maxima in Wils et al.\,(2010, 2011),
we draw here a $(O-C)$ diagram based on a total of 385 times of maximum light.

\begin{table}[th!!!]
\begin{center}
\caption{The new times of maximum light of V1162 Ori obtained in 2007--2011.}
  \label{maximum}
\begin{tabular}{cccccc}
\noalign{\smallskip}\hline
No.& HJD(max)    & No. &  HJD(max)     & No. &  HJD(max)  \\
\hline
1 & 2454130.0693 & 14 & 2454178.0626   & 27 & 2455525.13603 \\
2 & 2454131.0901 & 15 & 2454181.0580   & 28 & 2455525.21233 \\
3 & 2454131.1701 & 16 & 2454184.0422   & 29 & 2455525.29404 \\
4 & 2454132.1171 & 17 & 2454847.0331   & 30 & 2455526.15567 \\
5 & 2454132.1882 & 18 & 2454847.1111   & 31 & 2455526.23844 \\
6 & 2454133.0578 & 19 & 2454847.1889   & 32 & 2455526.31468 \\
7 & 2454133.1367 & 20 & 2454871.1100   & 33 & 2455565.10686 \\
8 & 2454133.2158 & 21 & 2454871.1910   & 34 & 2455565.18583 \\
9 & 2454137.0715 & 22 & 2455249.1147   & 35 & 2455565.26503 \\
10& 2454137.1467 & 23 & 2455250.0539   & 36 & 2455602.08888 \\
11& 2454161.0694 & 24 & 2455250.1342   & 37 & 2455602.16764 \\
12& 2454164.0582 & 25 & 2455524.19384  & 38 & 2455610.03529 \\
13& 2454176.0189 & 26 & 2455524.26824  & 39 & 2455610.11388 \\
\hline
\end{tabular}
\end{center}
\end{table}

\begin{figure*}[t!]
\centering
\includegraphics[angle=-90,width=0.90\columnwidth]{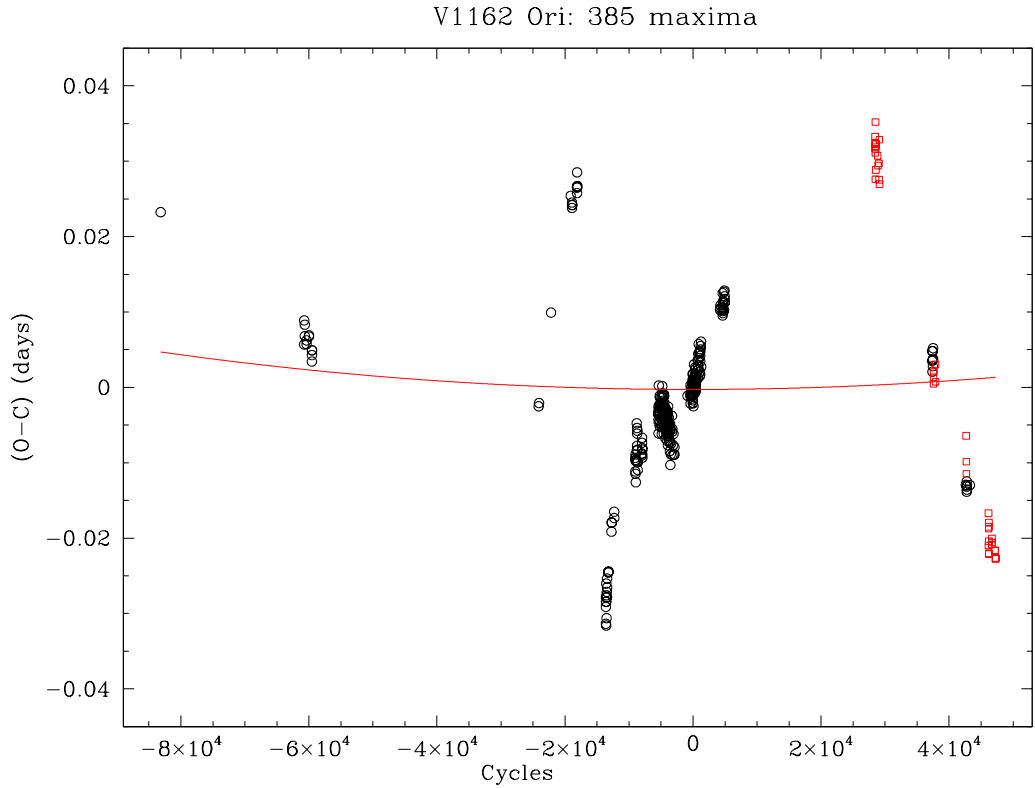}
\caption{
The $O-C$ diagram of V1162 Ori: the forced parabolic fitting curve is not justified.
The new maxima in Table 1 are drawn in squares.}
\label{fig:omc}
\end{figure*}

To calculate $(O-C)$ residuals and their corresponding cycles (denoted by {\em E} below)
elapsed since an initial maximum epoch,
we have defined a new ephemeris
\begin{equation}
T_{\rm max} = {\rm HJD}\, 2451890.3708 + 0.0786869 \times E .
\end{equation}
according to Arentoft \& Sterken (2002).
The cycle counts are usually the results rounded off by
rounding a numerical value to the nearest integer.
For those cases when a value is exactly half-way between two integers,
it is always rounded up following the so-called `round half up' tie-breaking rule.
For example, the values (66.51, 23.5, --23.5, 15.49, --18.38, --18.67),  get rounded to
(67, 24, --23, 15, --18, --19), respectively.
This rounding policy results in $(O-C)$ residuals within half a period,
i.e. $|O-C|\leq 0.0393$\,days.
The resultant $O-C$ diagram was drawn in Fig.~\ref{fig:omc}, which cannot be fitted by
either a single linear line or a simple parabolic curve.
Part of the 385 maxima and $(O-C)$ data are shown in Table~\ref{O-C}.
The full list is available upon request from the authors.

\begin{table}[h!!!]
\begin{center}
\caption{List of 385 light maxima of V1162 Ori and their $(O-C)$ residuals computed
by Eq.(1). The reference 'L' refers to Lampens(1985),
'A' refers to Arentoft et al.(2000,2001a,2001b,2002),
'H' stands for Hintz et al.(1998),
'P' for Poretti et al.(1990), 'W' for Wils et al.(2010,2011), while 'J' for this work.}
  \label{O-C}
\begin{tabular}{cccccc}
\noalign{\smallskip}\hline
\multicolumn{1}{c} {\textbf{No.}} &
\multicolumn{1}{c} {\textbf{Maximum}} &
\multicolumn{1}{c} {\textbf{Fractional Cycles}} &
\multicolumn{1}{c} {\textbf{Cycles}} &
\multicolumn{1}{c} {\textbf{$(O-C)$}} &
\multicolumn{1}{c} {\textbf{Ref.}} \\
\hline
   1 &   2445347.02750  &  --83156.70   &   --83157  &  0.023243  & L \\
   2 &   2447110.78000  &  --60741.88   &   --60742  &  0.008880  & P \\
   3 &   2447110.85550  &  --60740.92   &   --60741  &  0.005693  & P \\
   $\cdots\cdots$\\
 247 &   2451890.37080  &       0.00   &        0  &  0.000000  & A \\
 248 &   2451890.45000  &       1.00   &        1  &  0.000513  & A \\
    $\cdots\cdots$\\
 346 &   2455293.33020  &   43246.83   &    43247  & --0.012964  & W \\
 347 &   2454130.06930  &   28463.42   &    28463  &  0.033265  & J \\
    $\cdots\cdots$\\
 384 &   2455610.03529  &   47271.71   &    47272  & --0.022647  & J \\
 385 &   2455610.11388  &   47272.71   &    47273  & --0.022744  & J \\
\hline
\end{tabular}
\end{center}
\end{table}


\clearpage
\references
\newcommand{\ibvs}{{\em Inf. Bull. Variable Stars}}

Arentoft, T., Sterken, C., 2000, {\it A\&A}, {\bf 354}, 589

Arentoft, T., et al., 2001a, {\it A\&A}, {\bf 374}, 1056

Arentoft, T., et al., 2001b, {\it A\&A}, {\bf 378}, L33

Arentoft, T., Sterken, C., 2002, {\it ASP Conf. Ser.}, {\bf 256},
79, in Observational Aspects of Pulsating B- \& A Stars.,

Hintz, E. G., Joner, M. D., Kim, C., 1998, {\it PASP}, {\bf 110}, 689

Lampens, P., 1985, {\it IBVS}, 2794

Wils, P., Hambsch, F. J., Lampens, P., et al., 2010, {\it IBVS}, 5928

Wils, P., Hambsch, F. J., Robertson, C. W., et al., 2011, {\it IBVS}, 5977

Zhou, A.-Y., Liu, Z.-L., 2003, {\it AJ}, {\bf 126}, 2462
\endreferences


\end{document}